\documentclass[12 pt]{article}
\usepackage{graphicx}
\usepackage{amsmath}
\usepackage{epsfig}
\usepackage{textcomp}
\usepackage{amssymb}
\usepackage{stmaryrd}
\usepackage[usenames]{color}
\newcommand{\beq}{\begin{eqnarray}}
\newcommand{\eeq}{\end{eqnarray}}

\title{Pair Distribution Function of a Square-Well Fluid}
\author{Sanghamitra Neogi and Gerald D. Mahan \\
Department of Physics, Pennsylvania State University \\
University Park, Pennsylvania 16802, USA
}
\begin{document}
\maketitle
\begin{minipage}[t]{5in}
\begin{center}
Abstract
\end{center}
The properties of the ground state of liquid $^4$He are studied using a correlated basis function of the form $\prod_{i<j} \psi(r_{ij})$. Here, $\psi(r)$ is chosen as the exact solution of the Schr\"{o}dinger equation for two $^4$He atoms. A hard-sphere plus an attractive square well is used as the interaction potential between $^4$He atoms. The pair distribution function is calculated using approximate integral methods, namely the Percus-Yevick (PY) equation and Hypernetted Chain (HNC) approximation. The values thus obtained are used to calculate the ground state energy, which is found to be $-4.886$ K using the PY equation. The liquid structure factor is also obtained using the pair distribution function. The values for the pair distribution function and liquid structure factor are compared with experimental results and earlier theoretical calculations.
\end{minipage}

\section*{1. Introduction}
The ground state properties of rare gas fluids have been long studied (\cite{Likos} - \cite{Woodhead}). Many of these properties are determined by the knowledge of pair distribution function, $g(r)$, which can be measured experimentally. There have been many efforts to calculate the pair distribution function over the years. Most of the theoretical efforts employed methods mostly used in classical physics, such as Monte Carlo (\cite{Michler} - \cite{Ceperley4}). Alternately, the pair distribution function is determined by solving a nonlinear integral equation such as Percus-Yevick (PY) \cite{Percus_Yevick} or the Hypernetted Chain approximation (HNC) \cite{HNC}. Most of these calculations are done in the classical physics domain.

For liquid helium, quantum effects are important and has to be incorporated in the theory. Some of the calculations for obtaining the pair distribution function of liquid helium employed Quantum Monte Carlo methods (\cite{Ceperley1} - \cite{Ceperley3}) but most of the pair distribution function studies used only classical methods (\cite{McMillan} - \cite{Francis}). The properties of liquid helium can be understood to a great extent in terms of short-range correlations emanating from the strong interactions between particles. The interactions are usually represented by a central pair potential of short range, the most common form being the Lennard-Jones(LJ) potential. An alternate to the LJ potential is the hard-sphere potential. 

A major success of the hard-sphere potential was that the exact solution of the PY equation can be obtained for this potential (\cite{Thiele} - \cite{Lebowitz}). Several later works have extended the calculation by either modifying the hard-sphere potential (e.g. considering an extended hard core potential \cite{Lebowitz_Percus}) or adding another form of potential outside the hard-sphere (e.g. an attractive square-well potential). These calculations have been classical (\cite{Smith} - \cite{Dubinin}).

Here we present a quantum mechanical solution to $g(r)$ for a fluid with the hard-sphere potential and an attractive square well outside of it. The short-range correlations in liquid helium are commonly treated by writing the many-body wavefunction as a product of pair functions, known as the Jastrow function \cite{Jastrow}. We solve exactly the two-body Schrödinger equation for the hard-sphere plus attractive square well potential, and then use the pair wave function to construct the many-body Jastrow wavefunction. The important advantage of such a wavefunction is the formal analogy between its energy expectation values with the configuration-space integrals encountered in classical equilibrium statistical mechanics. We calculate pair distribution function, $g(r)$ of liquid $^4$He using two approximate integral equation methods, the Percus-Yevick (PY) and Hypernetted Chain (HNC) approximation. We use the values of the pair distribution function to obtain the ground state energy and liquid structure factor, $S(k)$. We compare our theoretically obtained values of $g(r)$ and $S(k)$ to the  experimental results (\cite{Gordon} - \cite{Svensson}) and also to earlier classical calculations (\cite{McMillan} - \cite{Francis}). Our theoretical results qualitatively matches the experimental data. However, one needs to use a more accurate form for the potential, as for example, the Aziz potential \cite{Aziz} to obtain a better quantitative agreement. One logical and very interesting extension of this work would be to investigate if the wavefunction obtained can describe supersolidity.

\section*{2. Theory}
\subsection*{2.1 Ground-State Wavefunction}

The Hamiltonian for a system of $N$ $^4$He atoms of mass $m$ in a volume $\Omega$, interacting through a central pair potential $V(|\vec{r}|)$ is given by,
\begin{equation}
H = - \sum_{i=1}^{N} \frac{\hbar^2 }{2m}\nabla_{i}^2 + \sum_{i<j} V(|\vec{r}_i-\vec{r}_j|). 
\label{H:1}
\end{equation}
The best wavefunction for liquid $^4$He is the correlated basis function of the Jastrow form. For a many-particle fluid, the wavefunction can be written as
\beq
\Psi(\vec{r_{1}},\vec{r_{2}}, \ldots, \vec{r_{N}}) &=& \prod_{i<j}^{N} \psi(|\vec{r}_{i} - \vec{r}_{j}|) \label{WF:1} \\
& = & \exp { \sum_{i<j} \ln [\psi(|\vec{r}_{i} - \vec{r}_{j}|)]} \label{WF:2}
\eeq
Now, we must choose a reasonable form for the pair function $\psi(r)$; this function should be small for short distances and should approach a constant for large distances. At small distances, where the two particles interact strongly the pair function is not expected to be very different from the solution of the two-body problem. If the potential function between pairs of $^4$He atoms has the form of a hard-sphere with an attractive square well,
\beq
V(r) = \left\{ \begin{array}{cc}
			\infty & r < a\\
		  -V_0 & a < r < b \\
			0 & b<r \ , \end{array} \right.
\label{V:1}
\eeq
then the Schr\"{o}dinger equation for the two atom wavefunction, in relative coordinates, is
\beq
E \psi(r) = \big[ -\frac{\hbar^2 \nabla^2}{m} + V(r) \big] \psi(r).
\eeq
The eigenfunctions $(l=0)$ can be written as
\beq
\psi(r) & = & \left\{ \begin{array}{cc}
			0 & r < a\\
		  A \sin[p(r-a)] & a < r < b \\
			\sin(kr+\delta) & b<r, \end{array} \right.
\label{WF:3} \\
p^2 &=& k^2 + \frac{m V_0}{\hbar^2}, \ E = \frac{\hbar^2 k^2}{m}. \label{p:1}
\eeq
Now, two $^4$He atoms form a very weakly bound state, with $E \approx 0$. Hence, we can approximate the eigenfunction as 
\beq
\psi(r) &=& \left\{ \begin{array}{cc}
			0 & r < a\\
		  \sin[p_0(r-a)] & a < r < b \\
			1 & b<r, \end{array} \right.
\label{WF:4} \\
p_0^2 & = & \frac{m V_0}{\hbar^2}, \label{p:2}\\
\sin[p_0(b-a)] & = & 1 ,\  p_0(b-a) = \frac{\pi}{2}.
\label{p:3}
\eeq

\subsection*{2.2. Integral Equation Methods}

The pair distribution function is defined as
\begin{equation}
g(\vec{R}_1, \vec{R}_2) = \Omega^2 \int \sum_{i \neq j} \delta(\vec{R}_1 - \vec{r}_i) \delta(\vec{R}_2 - \vec{r}_j) \Psi^2 d\vec{r}_1 \ldots d\vec{r}_N \bigg/ \Psi^2 d\vec{r}_1 \ldots d\vec{r}_N.
\label{g:1}   
\end{equation}
Since, we have a tranlslationally invariant system, the pair distribution function is a function of the relative coordinate, $(\vec{R}_1 - \vec{R}_2)$ only. In addition, $g(\vec{R}_1 - \vec{R}_2)$ is spherically symmetric in the liquid state, hence, $g(\vec{R}_1 - \vec{R}_2) = g(|\vec{R}_1 - \vec{R}_2|) = g(r)$. The pair distribution function is normalized in such a way that $g(r) = 1$ for large $r$ and it satisfies the sum rule,
\begin{equation}
n\int[1-g(\vec{r})]d\vec{r} = 1.
\label{g:2}
\end{equation}
One important advantage of using correlated basis functions is that the form of the diagonal density matrix $|\Psi|^2$ is mathematically identical to that of a classical fluid. The classical problem has been extensively studied, and accurate computational methods are available which work well for liquids of neutral atoms. The most successful methods are based on the Percus-Yevick(PY) integral equation and the Hypernetted Chain(HNC) approximation. The PY equation for the pair distribution function of a system of $^4$He atoms with pair wavefunction described by Eq.\ref{WF:4} is  
\begin{equation}
\frac{g(r)}{\psi^2(r)} = 1 + n \int d\vec{y} [g(\vec{y})-1] g(\vec{r}-\vec{y})[1 - \frac{1}{\psi^2(\vec{r}-\vec{y})}].
\label{PY:1}
\end{equation}
Performing the integration over angular variables and defining $\frac{g(r)}{\psi^{2}(r)} = g^\prime (r)$, Eq.\ref{PY:1} can be written as
\begin{equation}
g^\prime(r) = 1 + \frac{2\pi n}{r}  \int_0^{\infty} y dy [\psi^{2}(y) g^\prime (y) - 1] \int_{|r-y|}^{r+y} z dz g^\prime (z) [\psi^{2}(z) -1] .
\label{PY:2}
\end{equation}
On the other hand, the Hypernetted Chain integral equation (HNC) for the pair distribution function of the system of interest is given by  
\begin{equation}
\log \frac{g(r)}{\psi^{2}(r)} = n \int d\vec{y} [g(\vec{y})-1][g(\vec{r}-\vec{y}) -1 - \log \frac{g(\vec{r}-\vec{y})}{\psi^2(\vec{r}-\vec{y})}].
\label{HNC:1}
\end{equation}
Using a similar definition for $g^\prime (r)$ as in PY equation and after performing the integration over angular variables, we have
\begin{equation}
\log g^\prime(r)= \frac{2 \pi n}{r} \int^{\infty}_{0} y dy [\psi^{2}(y) g^\prime (y)-1] \int^{r+y}_{|r-y|} dz z [\psi^{2}(z) g^\prime (z)-1-\log g^\prime (z)]
\label{HNC:2}
\end{equation}
\subsection*{2.3. Ground State Properties}
The ground-state energy $E_0$ is given by the expectation values of the Hamiltonian in Eq.\ref{H:1},
\begin{equation}
E_0 = \langle H \rangle = \int \Psi H \Psi d\vec{r}_1 \ldots d\vec{r}_N \bigg/ \int \Psi^2 d\vec{r}_1 \ldots d\vec{r}_N. 
\label{E:1}
\end{equation}
If the many-body wavefunction, $\Psi$ has the form shown in Eq.\ref{WF:1}, it is straightforward to show that 
\begin{equation}
\int \Psi H \Psi d\vec{r}_1 \ldots d\vec{r}_N = \int \sum_{i<j} \big[ -\frac{\hbar^2 }{m}\nabla_{i}^2 \ln\psi(r_{ij}) + V(r_{ij}) \big] \Psi^2 d\vec{r}_1 \ldots d\vec{r}_N
\label{E:2}
\end{equation}
In terms of the pair distribution function, the potential energy per particle can be written as  
\begin{equation}
\frac{\langle P.E. \rangle}{N} =\frac{n}{2} \int d^{3} r V(r) g(r),
\label{V:2}
\end{equation}
and in our case, it is reduced to,
\begin{equation}
\frac{\langle P.E. \rangle}{N} = - 2 \pi n V_0 \int_a^b dr r^{2} g(r). 
\label{V:3} 
\end{equation}
The kinetic energy per particle can be obtained using
\begin{eqnarray}
\frac{\langle K.E. \rangle}{N} & = & \frac{n \hbar^{2}}{4 m} \int d^{3} r \frac{d (\ln [\psi(r)])}{dr} \frac{dg(r)}{dr}, 
\label{KE:1}
\end{eqnarray}
and, in our case, Eq.\ref{KE:1} becomes
\begin{equation}
\frac{\langle K.E. \rangle}{N} = \frac{n \hbar^{2} \pi }{m} ( p_0^2 \int_a^{b} dr g^{\prime} (r)r^2 - 2 p_0 \int_a^b dr g^{\prime}(r) r  \sin[p_0(r-a)] \cos[p_0(r-a)]). 
\end{equation}
The liquid structure function $S(k)$ (for $k \neq 0$) is related to the pair distribution function $g(r)$ by the following relation 
\beq
S(\vec{k}) &=& 1 +  n \int d \vec{r} [g(\vec{r})-1] \exp (-i \vec{k}.\vec{r}). 
\label{Sk:1}
\eeq
After carrying out the angular integration, the expression for liquid structure factor becomes
\begin{eqnarray}
S(\vec{k})&=& 1 + \frac{4 \pi n}{|\vec{k}|} \int_0^{\infty} r dr [g(r)-1] \sin(|\vec{k}||\vec{r}|). 
\end{eqnarray}

\section*{3. Method}
The PY integral equation (Eq.\ref{PY:2}) and the HNC approximation integral equation (Eq.\ref{HNC:2}) are solved self-consistently for $ r \leq R = 100$ \r{A}. The equilibrium density of the system of $^4$He atoms is chosen to be $n = 0.0218 \mbox{ \r{A}}^{-3}$ or $2.18 \times 10^{22}$ atoms/cm$^3$. The hard-sphere radius for the potential function is taken to be $a = 2.6$ \r{A}, the width and depth are chosen as $b = 4$ \r{A} and $V_0 = -15.26$ K respectively.  

In any numerical method it is necessary to truncate the infinite integrals at some stage. These integrals are replaced by a finite sum, assuming that $g(r) \approx 1$ beyond some large but finite radial distance $R$. We carried out two calculations for the self-consistent PY equation (Eq.\ref{PY:2}), one using $R = 30$ \r{A} and another $R = 100$ \r{A} to investigate the effect of the size of the cutoff. The maximum difference between the two sets of $g(r)$ values for all distances is $3e^{-5}$. For the HNC equations, we performed three calculations using $R = 50$ \r{A}, $60$ \r{A} and $70$ \r{A}. The three sets of $g(r)$ values differ not more than by $0.005$ for all distances. We used Simpson's three-point approximation to evaluate the integrals at every step of the iteration, using $N$ points evenly distributed over the range $(0,R)$. The step size chosen for the Simpson grid was 0.01, giving the value of $N$ as 10000 for $R = 100$ \r{A}. To investigate the influence of the step size on the accuracy of the results we performed two calculations for $R = 20$ \r{A}, one using 2000 points and one using 4000 points. 
It is found that the two sets differ nowhere by more than 0.002. The starting value of $g^\prime(r)$ is chosen to be 1 for all values of $r$. The input values for the next iteration is calculated according to the mixing formula
\begin{eqnarray}
g_{in}^{\prime (i)}(r) = \alpha g_{out}^{\prime (i-1)}(r) + (1-\alpha) g_{in}^{\prime (i-1)}(r). 
\end{eqnarray}
where $\alpha = 0.1$ is used to achieve desired convergence. The iterations are assumed to give convergence when value of the residual $Res$ is less than 1, where $Res$ defined by 
\begin{eqnarray}
Res = \sum{\big[g_{out}^{\prime (i)} - g_{out}^{\prime (i-1)}\big]^2}.
\end{eqnarray}

\section*{4. Results}
\begin{figure}
\centering
\includegraphics[width=5in]{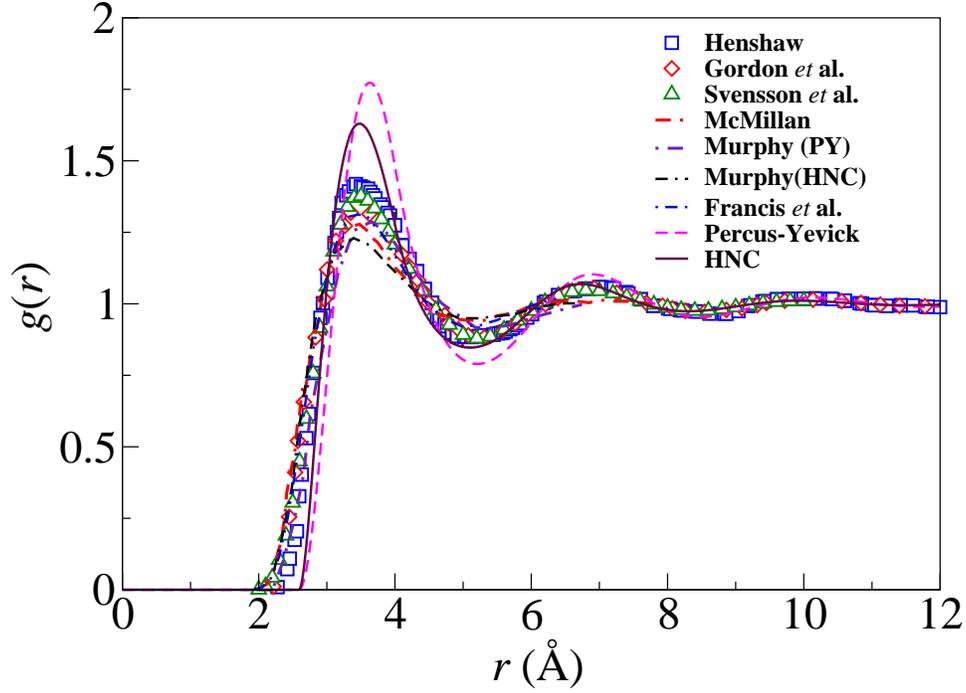}
\caption{(Color Online) Comparison of the computed quantum mechanical pair distribution function with earlier classical calculations and experimental data. The solid line represents the values obtained using Percus-Yevick equation, while the dashed line represents the values obtained using HNC equation. The diamonds are the x-ray data of Gordon \textit{et al.} \cite{Gordon}. 
The squares and the triangles show the neutron-diffraction data of Henshaw \cite{Henshaw} 
and Svensson \textit{et al.} \cite{Svensson} 
respectively. The dash-dash-dot (McMillan \cite{McMillan}), dash-dot (Murphy \cite{Murphy} using PY equation), dotted (Murphy \cite{Murphy} using HNC equation) and dash-dot-dot (Francis \cite{Francis}) lines represent earlier classical calculations.}
\label{Fig:1}
\end{figure}

The values of the pair distribution function for $^4$He obtained using PY (Eq.\ref{PY:2}) and HNC equations (Eq.\ref{HNC:2}) are shown in Fig.\ref{Fig:1}. The pair distribution function exhibits familiar features, e.g. $g(r)\rightarrow 1$ for large distances. The results satisfy the sum rule within numerical accuracy. Values obtained using the PY equation yield a sum of $0.939701$, while the values obtained using HNC equation give $0.811651$. Figure \ref{Fig:1} shows our results in comparison with earlier classical calcultaions (\cite{McMillan}-\cite{Francis}) as also the x-ray \cite{Gordon} and neutron-diffraction data \cite{Henshaw,Svensson}. The different symbols represent the experimental data while the different lines represent theoretical calculations.

We would like to point out here that while the earlier classical calculations yield values which are lower than the experimental data, our values for the pair distribution function lies above the experimental results. We would ascribe this discrepancy to the choice of parameters for the potential function. We used fixed values for the parameters $a, b, V_0$ throughout our calculation. Some of the earlier papers (classical calculations) \cite{Murphy,Dubinin1} have discussed the effect of varying the parameters to both the pair distribution function and the liquid structure factor. The earlier paper \cite{Murphy} has shown that the peak of the pair distribution function increases with the increase in the hard-sphere radius. And the recent article\cite{Dubinin1} has shown that the peak value of the liquid structure factor increases with the increase of width and depth of the square well part of the potential. Hence, we need to choose the values of the parameter carefully in order to obtain a good quantitative agreement between the theoretically obtained pair distribution function values and the experimental data. The better way would be to obtain the parameters using variational calculation.

The numerically obtained values of the pair distribution function is utilized to obtain the ground state energy for $^4$He. The average value for the potential energy per particle and the average value of the kinetic energy per particle are $-41.67$ K and $36.78$ K respectively. Hence, the average ground state energy per particle is found to be $-4.89$ K for the pair distribution function obtained using PY equation. This value differ from the experimentally obtained ground state energy of $^4$He atoms by about $31$\%. 

\begin{figure}\includegraphics[width=5in]{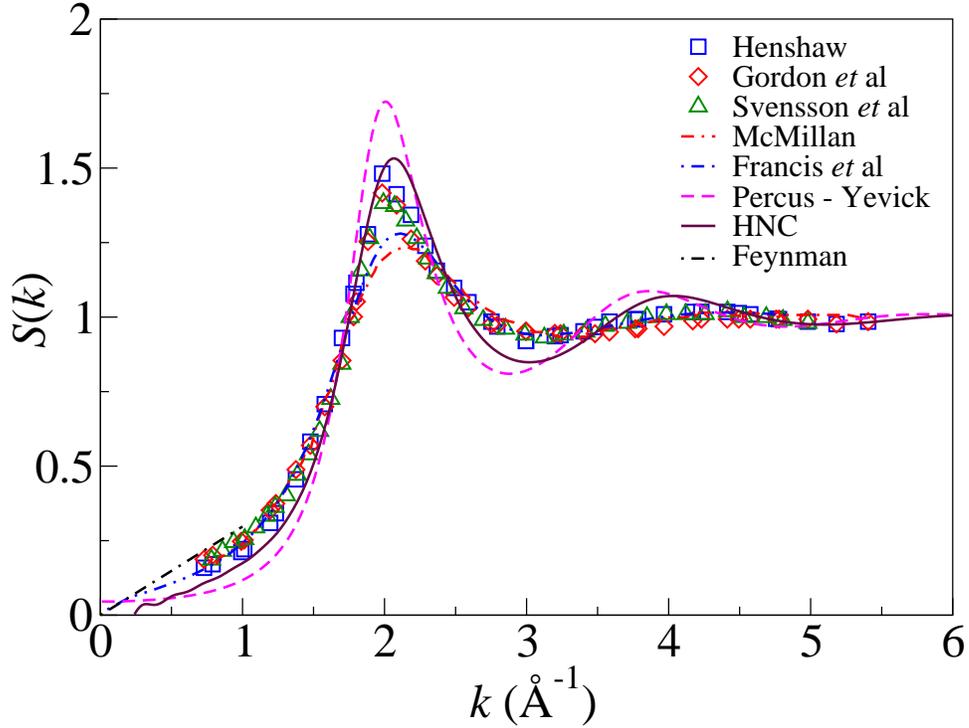}
\caption{(Color Online) Liquid structure factor values of $^4$He in comparison with experiment and earlier classical calculations. The solid line represents the values obtained using PY equation, while the dashed line represents the values obtained using HNC equation. Our results from HNC approximation match the experiment closely. The diamonds are the x-ray data of Gordon \textit{et al.} \cite{Gordon}. The squares and the triangles show the neutron-diffraction data of Henshaw \cite{Henshaw} and Svensson \textit{et al.} \cite{Svensson} respectively. The dash-dash-dot (McMillan \cite{McMillan}) and dash-dot-dot (Francis \cite{Francis}) lines represent earlier classical calculations. The dash-dot line is computed using Feynman \cite{Feynman} theory with the experimental velocity of sound (267 m/sec).}
\label{Fig:2}
\end{figure}

Using our results of the pair distribution function and the relation between the pair distribution function and the liquid structure factor(Eq.\ref{Sk:1}), we now calculate the liquid structure factor of $^4$He. The results are shown in Fig.\ref{Fig:2} along with earlier theoretical calculations as well as the x-ray and neutron-diffraction data. We use a similar representation as used in the graph for pair distribution function before: different symbols corresponds to the experimental data and different lines corresponds to theoretical calculations. The theoretical structure factor curves in Fig.\ref{Fig:2} agree well with the experimental data except in the region of diffraction maximum at $2$\r{A}$^{-1}$. Our results using the HNC approximation match the experimental results closely. Here again, we note that our calculation using PY equation produces a higher peak value while earlier classical calculations produce a lower peak compared to the experimental data. As we discussed in the second paragraph of this section, the choice of the values of the parameters in the potential function is responsible for this discrepancy between results from PY equation and experiment. 

The structure factor is expected to approach the Feynman value\cite{Feynman} for small $k$,
\begin{equation}
S(k) \approx \frac{\hbar k}{2mc},\ \ k\rightarrow 0,
\end{equation}
where $c$ is the velocity of sound. This limit is shown by the dash-dot line in Fig.\ref{Fig:2}. The structure factors obtained using both the PY and HNC equations approach a constant as $k$ tends to zero. The reason for this discrepancy is that we have restricted ourselves to a pair function that remains a constant for large distances. 

\section*{5. Discussion} 

We proposed a calculation of quantum mechanical pair distribution function using the Percus-Yevick and Hypernetted chain integral equations. The interaction potential between the $^4$He atoms are assumed to be given by a hard-sphere and an attractive square-well.    
There is a good qualitative agreement between the theoretically obtained values of the pair distribution function, the liquid structure factor and the experimental data. However, both of these computed physical quantities have higher peak values compared to the experimental results and the oscillations in the values are more pronounced for large distances. Also, the cutoff in the pair distribution function is shifted by a small amount. The parameters of the interaction potential adjust the position and sharpness of the cutoff and the peak in the pair distribution function. In order to get good quantitative agreement, we need to adjust the parameters of the interaction potential. It would be useful to incorporate a more realistic potential between the atoms, e.g. Aziz potential\cite{Aziz} and compare the pair distribution function, obtained using our proposed method, with experiments. 

\end{document}